\documentstyle[12pt]{article}
\begin{document}
\bibliographystyle{unsrt}
\vbox {\vspace{6mm}}

\begin{center}
{\bf DEFORMATIONS AND NONLINEAR SYSTEMS}
\end{center}

\begin{center}
V. I. MAN'KO,$^\star$\footnote{On leave from Lebedev Physical
Institute, Moscow, Russia.} G. MARMO,$^\dagger $ 
and F. ZACCARIA$^\dagger $
\end{center}

\begin{center}
$^\star ${\it Osservatorio Astronomico di Capodimonte\\
Via Moiariello 16, 80131 Napoli, Italy}
\end{center}  

\begin{center}
$^\dagger $ {\it Dipartimento di Scienze Fisiche,
Universit\`a di Napoli ``Federico II''\\
 and\\
Istituto Nazionale di Fisica Nucleare, Sezione di Napoli\\
Mostra d'Oltremare, Pad.19 - 80125 Napoli, Italy}
\end{center}

\begin{abstract}
The q--deformation of harmonic oscillators is shown to lead to
q--nonlinear vibrations. The examples of q--nonlinearized wave
equation and Schr\"odinger equation are considered. The procedure
is generalized to broader class of nonlinearities related to other 
types of deformations. The nonlinear noncanonical transforms used
in the deformation procedure are shown to preserve in some cases 
the linear dynamical equations, for instance, for the harmonic oscillators. The
nonlinear coherent states and some physical aspects of the 
deformations are reviewed.
\end{abstract}

\section{Introduction}

\noindent

We will review an approach to some nonlinear dynamical systems related 
to deformations of linear classical and quantum systems~\cite{sol1,bregenz}. 
The idea of the approach is to replace parameters of a linear system
with constants of the motion of a nonlinear system. This procedure produces
from a  linear system a nonlinear one and the q--oscillator 
of~\cite{bie,mc} may be considered as physical system with this specific 
nonlinearity. For q--oscillator, the constant parameter which was replaced by 
the constant of the motion depending on the amplitude of the vibration 
was the frequency. 

The approach may be used for many dynamical systems. So, the constant 
masses in Klein--Gordon and Dirac equations or the constant signal velocity 
in the wave equation may be replaced by  dynamical variables but 
these variables are chosen to be constants of the motion for the 
dynamical system under consideration. In fact, we address the problem 
whether physical constant parameters like light velocity, Planck constant, 
or gravitational constant are constant parameters in reality or they may 
depend, for example, on the initial condition in the process of evolution? 
The same question may be put about such characteristics of
elementary particle as electric charge or mass, say, of electron or fine
structure constant which play the role of constant parameters in the
dynamical equations of the theory. The corresponding nonlinear system 
of equations obtained in this approach may be simply a ``reparametrized''
initial linear system of equations in which constant parameters are
replaced by constants of the motion of the nonlinear system. 

One could deform also a simple initial nonlinear system containing some 
constant parameters and to transform it into another nonlinear system 
by replacing the constants with the integrals of the motion of the 
nonlinear system. So, it is possible to extend the class of integrable 
nonlinear systems starting from simple integrable nonlinear systems 
and reparametrizing them, i.e., replacing the constant parameters by 
the constants of the motion. 

The aim of our work is to give a review of the approach to q--deformation
based on the nonlinearization procedure and to present a general scheme 
for the described procedure of deforming the linear systems. 
We will consider examples of q--deformed harmonic oscillator, q-deformed
wave equation, and q--deformed Schr\"odinger equation.

The ansatz of deformation used in our approach is reduced to applying 
a nonlinear noncanonical transformation to oscillator complex amplitude.
This transformation adds to the complex amplitude a new factor which is 
integral of the motion for the oscillator motion both in linear and
nonlinear regimes of vibrations. Having thbility observation
will discuss,
about compatibility of system dynamics with different commutation
relations for the observables.

We will review the results of~\cite{sudar}, in which it was shown that one 
and the same harmonic oscillator dynamics is compatible with different
commutation relations of the oscillator quadratures. The nonlinear 
transformation of the oscillator amplitude by additing to it the 
factor which is constant of the motion leads naturally to the notion
of nonlinear coherent states~\cite{vogel} of f--coherent states~\cite{sud}.
Thus, we will review the properties of these states for which q--coherent
states~\cite{bie} are the partial case.

The physical consequences of the q--nonlinearity as blue shift 
effect~\cite{sol2}, deformation of Planck distribution 
formula~\cite{sol2,sol3}, and change of a charge form-factor~\cite{marman} 
will be discussed, as well as constructions of nonlinear coherent 
states~\cite{vogel} (f--coherent states~\cite{sud}\,).

\section{Quantum q--Oscillator}

\noindent 

The usual creation and annihilation oscillator operators $a$
and $a^\dagger$ obeying bosonic commutation relations 
\begin{equation}\label{qo1}
[a,a^\dagger]=1
\end{equation}
have in the Fock basis the known expressions 
\begin{equation}\label{qo2}
a=\left(\begin{array}{crcl}
0 & \sqrt{1} & 0 & \ldots \\
0 & 0 & \sqrt{2} & 0 \\
0 & 0 & 0 & \sqrt{3} \\
\ldots & \ldots & \ldots & \dots
\end{array}\right);~~~~~
a^\dagger=\left(\begin{array}{crcl}
0 & 0 & 0 & \ldots \\
\sqrt{1} & 0 & 0 & \ldots \\
0 &\sqrt{2} & 0 & \ldots \\
\ldots & \ldots & \ldots & \ldots 
\end{array}
\right).
\end{equation}
The q--oscillators may be introduced by generalizing the 
matrices~(\ref{qo2}) with the help of the q--integer numbers 
$n_{\mbox {q}}\,,$
\begin{equation}\label{qo3}
n_{\mbox {q}}=\frac{\sinh \,n\lambda }{\sinh \,\lambda }\,;
~~~\mbox {q}=e^{\lambda }\,.
\end{equation}
Here $\lambda $ and $\mbox {q}$ are dimensionless $c$--numbers. At 
$\lambda =0\,,\,\mbox {q}=1$ and the q--integer $n_{\mbox {q}}$ 
coincides with $n\,.$ 
Replacing the integers in~(\ref{qo2}) by q--integers we obtain  
matrices which define the annihilation and creation operators of the 
quantum q--oscillator,
\begin{equation}\label{qo4}
a_{\mbox {q}}=\left(\begin{array}{crcl}
0 &  \sqrt{1_{\mbox {q}}} & 0 & \ldots \\
0 & 0 & \sqrt{2_{\mbox {q}}} & \ldots \\
0 & 0 & 0& \sqrt{3_{\mbox {q}}}\\
\ldots & \ldots & \ldots & \ldots
\end{array}\right);~~~~~
a^\dagger _{\mbox {q}}=\left(\begin{array}{crcl}
0 & 0 & 0 & \ldots \\
\sqrt{1_{\mbox {q}}} & 0 & 0 & \ldots \\
0 & \sqrt{2_{\mbox {q}}} & 0 & \ldots \\
\ldots & \ldots & \ldots & \ldots
\end{array}\right).
\end{equation}
The above matrices obey the commutation relation
\begin{equation}\label{qo5}
[a_{\mbox {q}},\,a_{\mbox {q}}^\dagger]=F(\hat n)\,;
~~~\hat n=a^\dagger a\,,
\end{equation}
where the function $F(\hat n)$ has the form
\begin{equation}\label{qo6}
F(\hat n)=\frac {\sinh \,\lambda (\hat n+1)-\sinh \,\lambda \hat n}
{\sinh \,\lambda }\,.
\end{equation}
In addition, there exists the reordering relation
\begin{equation}\label{qo7}
a_{\mbox {q}}a_{\mbox {q}}^\dagger -{\mbox {q}}\,
a_{\mbox {q}}^\dagger a_{\mbox {q}}={\mbox {q}}^{-\hat n}\,.
\end{equation}
The operators $a_{\mbox {q}}$ and $a_{\mbox {q}}^\dagger $ can be 
expressed in terms of the operators $a$ and $a^ \dagger $
\begin{equation}\label{qo8}
a_{\mbox {q}}=af(\hat n)\,;~~~a_{\mbox {q}}^\dagger 
=f(\hat n)\,a^\dagger \,,
\end{equation}
where 
\begin{equation}\label{qo9}
f(\hat n)=\sqrt{\frac{a_{\mbox {q}}^\dagger a_{\mbox {q}}}
{a^\dagger a}}\,.
\end{equation}

The classical harmonic oscillator vibrating with unit frequency
may be described in terms of complex variables
\begin{equation}\label{qo10}
\alpha =\frac{q+ip}{\sqrt{2}}\,;~~~~~\alpha ^*=\frac{q-ip}{\sqrt{2}}\,,
\end{equation}
with nonzero Poisson bracket 
\begin{equation}\label{qo11}
\{\alpha ,\,\alpha ^*\}=-i\,.
\end{equation} 
Defining the classical q--oscillator in terms of new variables 
\begin{equation}\label{qo12}
\alpha _{\mbox {q}}=\sqrt{\frac{\sinh \,\lambda \alpha \alpha ^*}
{\alpha \alpha ^* \, \sinh \,\lambda }}\,\alpha \,; 
~~~~~\alpha ^*_{\mbox {q}}=\sqrt{\frac{\sinh \,\lambda \alpha \alpha ^*}
{\alpha \alpha ^*\,\sinh \,\lambda }}\,\alpha ^*\,,
\end{equation}
we get the Poisson brackets
\begin{equation}\label{qo13}
\{\alpha _{\mbox {q}},\,\alpha ^*_{\mbox {q}}\}
=-i\,\frac{\lambda }{\sinh \,\lambda }
\,\sqrt{1 +|\alpha _{\mbox {q}}|^4\,(\sinh \,\lambda )^2}\,. 
\end{equation}
We will consider a new system described by such q--variables with the
Hamiltonian function
\begin{equation}\label{qo14}
H(\alpha _{\mbox {q}},\,\alpha ^*_{\mbox {q}})
=\alpha _{\mbox {q}}\alpha _{\mbox {q}}^*\,. 
\end{equation}
Then the equations of the motion are 
\begin{equation}\label{qo15}
\dot \alpha _{\mbox {q}}=-i\,\frac {\lambda }{\sinh \,\lambda }\,
\sqrt {1+|\alpha _{\mbox {q}}|^4(\sinh \,\lambda )^2\,}
\,\alpha _{\mbox {q}}
\end{equation}
with solutions
\begin{equation}\label{qo16}
\alpha _{\mbox {q}}(t)=\alpha _{\mbox {q}}(0)\,\exp 
\left [\,\frac {-it\lambda }{\sinh \,\lambda }
\,\sqrt {1+|\alpha _{\mbox {q}}(0)|^4(\sinh \,\lambda )^2\,}\,\right ].
\end{equation}
We have performed a noncanonical transformation, i.e., deformed the Poisson
bracket, while preserving the form of the Hamiltonian.
Such a system can be rewritten in terms of $(\alpha , \alpha ^*)$
variables: in these coordinates, the original Poisson bracket is unchanged
while the Hamiltonian is  
\begin{equation}\label{qo17}
H_{\mbox {q}}(\alpha , \alpha ^*)=\frac {\sinh \,\lambda \alpha \alpha ^*}
{\sinh \,\lambda }\,.
\end{equation}
This dynamical system has a phase portrait which is the same 
as the usual linear harmonic oscillator. The equations of the motion for 
the system with the Hamiltonian $H_{\mbox {q}}$ are
\begin{equation}\label{qo18}
\dot \alpha =-i\omega _{\mbox {q}}\alpha \,;~~~~~
\dot \alpha ^*=i\omega _{\mbox {q}}\alpha ^*
\end{equation}
with
\begin{equation}\label{qo19}
\omega _{\mbox {q}}\equiv \omega _{\mbox {q}}(\alpha \alpha ^*)
=\frac{\lambda }{\sinh \,\lambda }\,
\cosh \,\lambda \alpha \alpha ^*\,.
\end{equation}
We notice that $\alpha \alpha ^*$ is a constant of the motion for 
the system, the frequency of which depends on the orbit.

This leads to interpreting q--oscillators as systems carrying a particular 
nonlinearity. What in the harmonic oscillator is a constant frequency
characterizing the evolution along any orbit, becomes a function 
constant on each orbit, separately. For the q--oscillator, the frequency 
depends on amplitude of vibrations. 

The nonlinear second order equation for the coordinate of the
oscillator $q$ is
\begin{equation}\label{qo20}
\ddot q+\omega _{\mbox {q}}^2q=0\,.
\end{equation}
It means that the dynamics of the coordinate $q$ and momentum $p$ of 
the nonlinear q--oscillator is described by the system of equations
\begin{equation}\label{qo21}
\dot q=\omega _{\mbox {q}}(\alpha \alpha ^*)\,p\,;~~~~~
\dot p=-\omega _{\mbox {q}}(\alpha \alpha ^*)\,q\,,
\end{equation}
where
\begin{equation}\label{qo22}
\alpha \alpha ^*=\frac {1}{2}\,(q^2+p^2)\,.
\end{equation}
For the deformed equations of the motion of the q--oscillator, in view of 
(\ref{qo21}), the momentum is the function of the velocity and position. 
This function may be obtained as the solution to the functional equation
\begin{equation}\label{qo23}
p\,(q,\,\dot q)=\frac {\sinh \,\lambda }{\lambda }\,\frac {\dot q}
{\cosh \,\{(\lambda /2)[q^2+p^2(q,\,\dot q)]\}}\,,
\end{equation}
considered as the implicit formula for the giving momentum as the function
of the position $q$ and velocity $\dot q\,.$ The solution to the
q--oscillator equation of the motion
\begin{equation}\label{qo24}
\dot \alpha =-i\alpha \,\frac {\lambda }{\sinh \,\lambda }\,\cosh \,\lambda
\alpha \alpha ^*
\end{equation}
is
\begin{equation}\label{qo25}
\alpha \,(t)=\alpha _0\exp \left [-it\,\frac {\lambda }{\sinh \,\lambda }
\,\cosh \lambda \alpha _0\alpha _0^*\right ],
\end{equation}
where 
$$\alpha _0=\alpha \,(t=0)$$
is the initial complex amplitude of the nonlinear q--oscillator. The 
solution to the equation of the motion for the coordinate $q$ of the 
nonlinear q--oscillator 
\begin{equation}\label{qo26}
\ddot q+\frac {\lambda ^2}{\sinh ^2\lambda }\,
\cosh ^2\left \{\frac {\lambda }{2}\,[q^2+p^2(q,\,\dot q)]\right \}=0\,,
\end{equation}
where the function $p\,(q,\,\dot q)$ is given implicitly by 
relation~(\ref{qo23}), may be written in the form
\begin{eqnarray}\label{qo27}
q\,(t)&=&\frac {q_0}{2}\left \{\exp \left [\frac {i\lambda t}
{\sinh \,\lambda }\,\cosh \left \{\frac {\lambda }{2}\left [q_0^2
+p^2(q_0,\,\dot q_0)\right ]\right \}\right ]\right.\nonumber\\
&+&\left.\exp \left [-\frac {i\lambda t}{\sinh \,\lambda }\,
\cosh \left \{\frac {\lambda }{2}\left [q_0^2
+p^2(q_0,\,\dot q_0)\right ]\right \}\right ]\right \}\nonumber\\
&+&\frac {\dot q_0\sinh \,\lambda }{2\,i\lambda }\,\cosh ^{-1}
\left \{\frac {\lambda }{2}\left [q_0^2
+p^2(q_0,\,\dot q_0)\right ]\right \}\nonumber\\
&\times &\left \{\exp \left [\frac {i\lambda t}{\sinh \lambda }
\,\cosh \left \{\frac {\lambda }{2}\left [q_0^2
+p^2(q_0,\,\dot q_0)\right ]\right \}\right ]\right.\nonumber\\
&-&\left.\exp \left [-\frac {i\lambda t}{\sinh \,\lambda }
\,\cosh \left \{\frac {\lambda }{2}\left [q_0^2
+p^2(q_0,\,\dot q_0)\right ]\right \}\right ]\right \}.
\end{eqnarray}
Here $q_0=q\,(t=0)$ and $\dot q_0=\dot q\,(t=0)$ are the initial 
position and velocity of the nonlinear q--oscillator. In the limit 
$\lambda \rightarrow 0\,,$ we have the standard solution for the linear 
harmonic oscillator.

One can find for small nonlinearity $\lambda \ll 1$ the approximate 
expression for the momentum solving Eq.~(\ref{qo23}) by iteration method. 
We have
\begin{equation}\label{qo28}
p=\dot q\left [1+\frac {\lambda ^2}{6}-\frac {\lambda ^2}{8}\left (q^2
+\dot q^2\right )\right ].
\end{equation}
Formula~(\ref{qo28}) may be interpreted as the negative shift of 
the oscillator mass by the factor depending quadratically on the energy 
of the oscillations.

\section{Deformed Wave Equations}

\noindent

Following~\cite{bregenz} we start from the wave equation of the form
\begin{equation}\label{dwe1}
\left (\frac{\partial ^2}{\partial t^2}-
\frac{\partial ^2}{\partial x^2}\right )\,\varphi \,(x,t)=0
\end{equation}
and represent this equation as a system of equations for decoupled 
oscillators. To do this, we rewrite Eq.~(\ref{dwe1}) in momentum 
representation
\begin{equation}\label{dwe2}
\ddot \varphi \,(k,t)+k^2\varphi \,(k,t)=0\,,
\end{equation}
where the complex Fourier amplitude
\begin{equation}\label{dwe3}
\varphi \,(k,t)=\frac{1}{2\,\pi }\,\int \varphi \,(x,t)\,\exp \,(-ikx)~dx
\end{equation}
plays the role of new coordinate. Since $\varphi \,(x,\,t)
=\varphi ^*(x,\,t)\,,$ we have
$\varphi \,(k,\,t)=\varphi ^*(-k,\,t)\,.$
Equation~(\ref{dwe2}) describes a two-dimensional oscillator with equal 
frequencies for both modes labeled by $k$ and $-k\,.$ Writing 
Eq.~(\ref{dwe2}) in the form
\begin{equation}\label{dwe4}
\dot \varphi \,(k,\,t)=\pi \,(k,\,t)\,;~~~~~
\dot \pi \,(k,\,t)=-k^2\varphi \,(k,\,t)\,,
\end{equation}
we have the equations in phase spaces of the oscillators with
nonunit frequency 
$\omega ^2=k^2,~~p\rightarrow \pi \,(k,\,t)\,,$ and 
$q\rightarrow \varphi \,(k,\,t)\,.$ Due to this, we can deform this 
linear system taking the integral of the motion 
\begin{equation}\label{dwe5}
\mu=\int dk\,\left \{\frac {1}{2\,|k|}\left [k^2|\varphi |^2(k,\,t)
+|F_k|^2\right ]\right \},
\end{equation}
in which the function $F_k$ playing the role of complex momentum of 
$k$--field mode is a solution to the infinite system of equations
\begin{equation}\label{dwe6}
\dot \varphi \,(k,\,t)=F_kf_{\mbox {q}}\left \{\int dk'
~\frac {1}{2\,|k'|}
\left [k^{'2}|\varphi |^2(k',\,t)+|F_{k'}|^2\right ]\right \}.
\end{equation}
The function $f_{\mbox {q}}$ is 
\begin{equation}\label{dwe7}
f_{\mbox {q}}(z)=\frac {\lambda }{\sinh \,\lambda }\,\cosh \,(\lambda z)\,.
\end{equation}
The parameter $\mu $ plays the role of the initial number of vibrations 
corresponding to given Cauchy initial conditions. 
Then the deformed wave equation may be written in the form
\begin{equation}\label{dwe8}
\ddot \varphi \,(x,\,t)=f_{\mbox {q}}^2(\mu )\,\frac {\partial ^2}
{\partial  x^2}\,\varphi \,(x,\,t)\,.
\end{equation}
We have the differential-functional equation which looks like standard
wave equation with the wave velocity $f_{\mbox {q}}(\mu )$ being a 
constant of the motion. Thus, the procedure of deformation yields us 
the nonlinear equation for which the velocity of wave propagation depends 
on the initial configuration of the field and its time derivative. 

The q--deformed wave equation (\ref{dwe8}) has the soliton-like solutions
\begin{equation}\label{dwe9}
\varphi _{\pm }(x,\,t)=\Phi \left (x\pm f_{\mbox {q}}(\mu )\,t\right ),
\end{equation}
where $\Phi $ is an arbitrary function. In fact, discussed 
q--deformation implies the existence of nonlinear interaction
among the modes. The q--deformed Klein--Gordon equation is considered by 
this method in~\cite{zac}. Some deformed relativistic equations are 
discussed in~\cite{luk,pil} from different viewpoints.

\section{Nonlinear Quantum Equation for One-Level System}

\noindent

Following~\cite{sudar} we start with a one-level quantum system to show 
that the Schr\"odinger equation for it gives rise to a Hamiltonian 
dynamics for a one-dimensional harmonic oscillator. In fact, 
the Schr\"odinger equation 
\begin{equation}\label{a1}
i\hbar \,\frac {\partial \Psi (t)}{\partial t}=H\Psi (t)\,,
\end{equation}
where $\Psi (t)$ is the wave function of the one-level system, is 
described by the Hamiltonian $H\,.$ This Hamiltonian is the Hermitian
1$\times $1--matrix and it means that the Hamiltonian is simply a 
$c$--number which is real. 

If one introduces the two real variables 
$q\,(t)$ and $p\,(t)$ as real and imaginary parts of the wave function
\begin{equation}\label{a2}
\Psi (t)=\frac {1}{\sqrt 2}\left [\frac {H}{\hbar }\,q\,(t)
+i\,p\,(t)\right ]\,,
\end{equation}
the Schr\"odinger equation aquires the form
\begin{equation}\label{a3}
\dot q=p\,;~~~~~
\dot p=-\frac {H^2}{\hbar ^2}\,q\,.
\end{equation}
Then introducing the frequency $\omega =H/\hbar \,,$
one can rewrite Eqs.~(\ref{a3}) as
\begin{equation}\label{a5}
\ddot q+\omega ^2q=0\,,
\end{equation}
which represents the equation of the motion for the one-mode harmonic
oscillator with the Hamiltonian
\begin{equation}\label{a6}
H=\frac {p^2}{2}+\frac {\omega ^2q^2}{2}\,.
\end{equation}
We may call this system  a classical-like system (the observation that 
quantum wave equation may be rewritten as classical-like equation was done
in~\cite{str}). 

The procedure used in Section~2 gives possibility to 
nonlinearize the linear Schr\"odinger equation. For one-level system,
time units can be chosen in such a way that the frequency $\omega $ is
equal to unity. Then, for example, the method of q--nonlinearization
of Section~2 gives 
\begin{equation}\label{a7}
i\dot \Psi =\left [\frac {\lambda }{\sinh \,\lambda }\,\cosh \,(\lambda
|\Psi |^2)\right ]\Psi \,.
\end{equation}
Equation~(\ref{a7}) is a nonlinear Schr\"odinger equation obtained in the 
frame of using q--nonlinearity of vibration of the complex probability 
amplitude. The procedure may be easily extended for multilevel quantum
system. We simply apply the same procedure of the q--nonlinearization
to the stationary state wave functions belonging to chosen energy levels,
because real and imaginary parts of the stationary wave functions are just
coordinates and momenta of the one-dimensional oscillators.

\section{Quantum Oscillations and Commutation Relations}

\noindent

In~\cite{wig}, Wigner studied the following problem: To what extent do the
equations of the motion determine the quantum mechanical commutation relations?
He found that the commutation relations are not uniquely determined  by
the equations of the motion even if the form of the Hamiltonian is fixed.
The problem was discussed in detail in~\cite{sudar} and below we review
the results of this work.

The equation of the motion for the linear quantum oscillator, with the
frequency and mass such that $\omega =m=1$ and Planck constant is
equal to unity, considered in~\cite{wig} may be written in terms 
of complex amplitude operators $a$ (annihilation operator) and $a^\dagger $
(creation operator) as
\begin{equation} \label{x1}
\dot a+ia=0\,;~~~~~
\dot a^\dagger -ia^\dagger =0\,.
\end{equation}
The commutation relation~(\ref{qo1}) may be chosen for these two operators. 
The Hamiltonian describing the linear equation~(\ref{x1}) may be taken in 
standard form
\begin{equation}\label{x3}
H=a^\dagger a+\frac {1}{2}\,.
\end{equation}
Using Eq.~(\ref{x3}) one can check that the Heisenberg equation of 
the motion for the complex amplitude operator $a$
\begin{equation}\label{x4}
i\dot a=[a,H]
\end{equation}
yields Eq.~(\ref{x1}). 

Now we will show that there exists another alternative description pair,
Hamiltonian--commutation relation, which produces the same dynamics. To do 
this, we first multiply Eq.~(\ref{x1}) from the right-hand side by the 
operator $f(\hat n)\,,$ where $\hat n=a^\dagger a$ and the real function 
$f(x)$ has no zeros at nonnegative integers. Since the operator $\hat n$ 
is the integral of the motion for the dynamics given by Eq.~(\ref{x1}), 
a function of the integral of the motion is also constant of the motion. 
Then we have for the operator
\begin{equation}\label{x5}
A=af(\hat n)
\end{equation}
the evolution equation 
\begin{equation}\label{x6}
\dot A+iA=0\,.
\end{equation}
The nonlinear transformation~(\ref{x5}) preserves the linearity of the 
particular equation~(\ref{x1}). Such transformations may be called 
{\it linearoid} in analogy with {\it canonoid} in classical mechanics, 
which are noncanonical transformations in general but, for particular 
case, preserve the symplectic structure (Poisson brackets). The 
linearoid~(\ref{x5}) has inverse
\begin{equation}\label{x7}
a=A\,\frac {1}{f(\hat n)}\,,
\end{equation}
in which we have to use 
\begin{equation}\label{x8}
A^\dagger =f(\hat n)\,a^\dagger
\end{equation}
and
\begin{equation}\label{x9}
\hat N=A^\dagger A=f^2(\hat n)\,\hat n= F(\hat n)
\end{equation}
to express the operator $\hat n$ as the function of the operator $N$
\begin{equation}\label{x10}
\hat n= F^{-1}(\hat N)\,.
\end{equation}
The function $y= F^{-1}(x)$ is the inverse function of the relation 
$x= F(y)\,.$ Relation~(\ref{x7}) may be rewritten as 
\begin{equation}\label{x11}
a=A\,\frac{1}{f\left [ F^{-1}(\hat N)\right ]}
\end{equation}
and also
\begin{equation}\label{x12}
a^\dagger =\frac{1}{f\left [ F^{-1}(\hat N)\right ]}\,A^\dagger \,.
\end{equation}
The operators $A$ and $A^\dagger $ satisfy the commutation relations
\begin{equation}\label{x13}
AA^\dagger-A^\dagger A=\varphi \left [ F^{-1}(\hat N)\right ]\,,
\end{equation}
where the function $\varphi \,(z)$ is related to the function $f(z)$ 
from Eq.~(\ref{x5}) by 
\begin{equation}\label{x14}
\varphi \,(z)=(z+1)\,f^2(z+1)-z\,f^2(z)\,.
\end{equation}
It is obvious, that since we performed only a change of variables using 
the invertible nonlinear noncanonical transformation (\ref{x5}), the 
Hamiltonian for the variable $A$ obeying the commutation 
relations~(\ref{x13}) and the linear equation of the motion~(\ref{x6}) 
is the same Hamiltonian~(\ref{x3}) but expressed in terms of the variables 
$A$ and $A^\dagger \,,$ i.e.,
\begin{equation}\label{x15}
H=\widehat F^{-1}(N)+\frac {1}{2}\,.
\end{equation}
On the other hand, we know that we could define (with another Hilbert 
space structure) the operators $B$ and $B^\dagger $ with commutation 
relations
\begin{equation}\label{x16}
BB^\dagger -B^\dagger B=1
\end{equation}
and with the Hamiltonian
\begin{equation}\label{x17}
H^{\prime}=B^\dagger B+\frac {1}{2}\,,
\end{equation}
obeying to the Heisenberg equations of the motion
\begin{equation}\label{x18}
\dot B+iB=0~~~~~~~~{\dot B}^\dagger - i B^\dagger = 0 \,.
\end{equation}
Thus, we conclude that for one and the same dynamics, i.e., one and the
same linear equations of the motion~(\ref{x6}) and (\ref{x18}), there exists
an infinite number of Hamiltonians and commutation relations giving the same 
dynamics. The ambiguity in choosing alternative Hamiltonians is labeled in 
the demonstrated procedure by a function $f$ of the oscillator energy.

It is worthy to note that the commutation relation~(\ref{x13}) implies
that the dispersions of quadratures
$$
P=\frac {1}{i\,\sqrt 2}\,(A-A^\dagger )~~~~~\mbox {and}~~~~~
Q=\frac {1}{\sqrt 2}\,(A+A^\dagger )$$
do not satisfy the Heisenberg uncertainty relation. 
But according to the dynamical equation~(\ref{x6}) these quadratures 
evolve performing usual harmonic oscillations. 

It is known~\cite{marmo,dodma}, that there exists an infinite number of 
variational formulations for one-dimensional classical systems. It is 
possible to use this alternative Hamiltonian descriptions to construct 
alternative quantum descriptions for the oscillator~\cite{sudar}. 

\section{One-Mode f--Coherent States}

\noindent

Coherent states were originally introduced as eigenstates of the 
annihilation operator for the harmonic oscillator and then widely 
used in physics, particularly in quantum optics. This is therefore 
a concept of algebraic origin and having now constructed a similar 
annihilation operator it is natural, following the same procedure, 
to construct a new class of f--coherent states (nonlinear coherent
states )in the Fock space. 
Further the f--coherent states may not be preserved under time 
evolution. Nevertheless, we are willing to call them
``f--coherent states'' for an easy identification, of the kind already 
proposed for the eigenstates of the q--annihilation operator which 
were named ``q--coherent states''~\cite{bie}.

Let us take for the one-mode case the operator~(\ref{x5}).
Then one can consider the eigenfunctions of 
$A,\mid \alpha , f\rangle $ in a Hilbert space. They therefore satisfy 
the equation
\begin{equation}\label{FC3}
A\mid \alpha, f\rangle =\alpha \mid \alpha ,f\rangle , ~~~~~\alpha 
\in \hbox {\bf C}\,.
\end{equation}
Looking for the decomposition of $\mid \alpha ,f\rangle $ in the Fock space
\begin{equation}\label{FC4}
\mid \alpha ,f\rangle =\sum _{n=0}^\infty c_n\mid n\rangle ,
\end{equation}
where $\mid n\rangle ,$ eigenfunction of $\hat n\,,$
is a normalized Fock state, we obtain
\begin{equation}\label{FC7}
c_n=N_{f,\,\alpha }\frac {\alpha ^n}{\sqrt {n!\,}\,[f(n)]!\,}\,,
\end{equation}
in which
\begin{equation}\label{FC8}
[f(n)]!=f(0)f(1)\cdots f(n)
\end{equation}
and
\begin{equation}\label{FC10}
N_{f,\,\alpha }=\left(\sum_{n=0}^\infty 
\frac {|\alpha|^{2n}}{n!\,|[f(n)]!|^2}\right)^{-1/2}.
\end{equation}
The scalar product is easily written
\begin{equation}\label{FC13}
\langle \alpha ,f\mid \beta ,f\rangle =N_{f,\,\alpha }\,N_{f,\,\beta } 
\,\sum _{n=0}^\infty\,\frac {1}
{n!\,|[f(n)]!|^2}\,(\alpha ^*\beta )^n\,.
\end{equation}
It should be remarked furthermore that, given $C(n)=C_n$
any real function on $\hbox {\bf Z}^+,$ the state 
$\mid \alpha ,C\rangle $ 
defined by
\begin{equation}\label{FC14}
\mid \alpha ,C\rangle =\sum _{n=0}^\infty C_n\,\alpha ^n\mid n\rangle
\end{equation}
is an eigenfunction of some $A\,.$ In fact, the corresponding
function $f$ is found to be
\begin{equation}\label{FC15}
f(n)=\frac {1}{\sqrt n}\,\frac {C_{n-1}}{C_{n}}\,. 
\end{equation}
In case $f(n)=1\,,$ we have standard coherent states.

The known q--coherent states~\cite{bie} turn out to be a particular 
case of f--coherent states. The normalization factor of such states is
\begin{equation}\label{FQ18}
N_{q,\,\alpha }
=\left ( \sum _{n=0}^{\infty }\frac {|\alpha|^{2n}}{[n]!}\right)^{-1/2},
\end{equation}
in which
\begin{equation}\label{FQ19}
[n]!=\frac {\sinh \,\lambda n}{\sinh \,\lambda}\,
\frac {\sinh \,\lambda (n-1)}{\sinh \,\lambda}\,\cdots \,1\,.
\end{equation}
For the scalar product, we have
\begin{equation}\label{FQ21}
\langle \alpha \mid \beta \rangle _\lambda 
=\left (\sum _{n=0}^\infty \frac {|\alpha |^{2n}}{[n]!}\right )^{-1/2}\,
\left (\sum _{n=0}^\infty \frac {|\beta |^{2n}}{[n]!}\right )^{-1/2}\,
\sum _{n=0}^\infty \frac {1}{[n]!}\,(\alpha ^*\beta )^n.
\end{equation}

\section{Deformation of Planck formula}

\noindent

We will discuss what physical consequences may be found if the 
considered f--nonlinearity influences the vibrations of the real 
field mode oscillators like, for example, electromagnetic field 
oscillators or the oscillations of the nuclei in polyatomic molecules.

First it will be seen that this nonlinearity changes the specific 
heat behaviour. To show this, we have to find the partition function 
for a single f--oscillator corresponding to the Hamiltonian 
$H=(AA^\dagger +A^\dagger A)/2\,,$ which is
\begin{equation}\label{1dp1}
Z(T)=\sum_{n=0}^{n=\infty}\exp \,(-\beta E_{n})\,,
\end{equation}
where the variable $\beta $ is the inverse temperature $T^{-1}.$ 
Calculating the partition function for an ensemble of 
q--oscillators~\cite{sol1}, we obtain that the 
specific heat decreases for $T\rightarrow \infty $ as
\begin{equation}\label{1dp2}
C\propto \frac {1}{\ln \,T}\,.
\end{equation}
Thus, the behaviour of the specific heat of the q--oscillator
is different from the behaviour of the usual oscillator 
in the high temperature limit. This property may
serve as an experimental check of the existence of vibrational
nonlinearity of the q--oscillator fields.

q--deformed Bose distribution can be obtained by the same
method and one has for small q--nonlinearity parameter
the following q--deformed Planck distribution formula~\cite{sol3} 
\begin{equation}\label{1dp5}
\langle n\rangle =\frac {1}{e^{\hbar \omega /kT}-1}
-\lambda ^2\,\frac {\hbar \omega}{kT}\,\frac {e^{3\hbar \omega /kT}
+4\,e^{2\hbar \omega /kT}+e^{\hbar \omega /kT}}
{(e^{\hbar \omega /kT}-1)^{4}}\,.
\end{equation}
It means that q--nonlinearity deforms the formula for the mean photon
numbers in black body radiation~\cite{sol1}. 

For small temperature, the behaviour of the deformed Planck 
distribution differs from the usual one~\cite{sol3}           
\begin{equation}\label{1dp6}
\langle n\rangle -\bar n_0=-\lambda ^2\,\frac {\hbar \omega }{kT}
\,e^{-\hbar \omega /kT}.
\end{equation}
As it was suggested in~\cite{sol2}, the q--nonlinearity of the field
vibrations produces blue shift effect which is the effect of the
frequency increase with the field intensity. For small nonlinearity
parameter $\lambda $ and for large number of photons $n$ in
a given mode, the relative shift of the light frequency is
$$\frac {\delta \omega }{\omega }=\frac {\lambda ^2n^2}{2}\,.$$
This consequence of the possible existence of a q--nonlinearity may
be relevant for models of the early stage of the Universe.

Another possible phenomenon related to the q--nonlinearity was
considered in~\cite{marman} where it was shown that if one deforms
the electrostatics equation using the method of deformed creation
and annihilation operators a point charge acquires a formfactor
due to q--nonlinearity.

\section{Conclusions}

\noindent

Starting with the example of the harmonic oscillator we have
exhibited a family of associated nonlinear system which are
completely integrable, both in classical and quantum physics.

We have shown that q--nonlinearity, associated with the  
Heisenberg--Weyl quantum groups, is a subcase of a more general class 
of possible nonlinearity. 

A class of states has been considered in the Fock space through
the deformation process applied to the harmonic oscillator operators.
Such states have been described as f--coherent states, or
nonlinear coherent states, and q--coherent states being particular 
examples of them. 

The studied nonlinearities, if they exist, for the electromagnetic
field or for the gluons, may influence the particle decays,
correlations in particle multiplicities.

The deformation of the standard oscillator was interpreted as describing 
the behaviour of a nonlinear oscillator for which the frequency of
vibrations depends on the amplitude of the vibrations. There is another
physical aspect of this problem related to the change of quadratute 
commutation relations which, in fact, incorporate the influence of
the nonlinearity of vibrations. If the oscillations of the quadratures
are oscillations of a field, the change of the commutation relations
means the different statistics of the field.
Thus, from the introduced interpretation of the deformed 
oscillators~\cite{sol1,sol2}, it follows that the statistical properties 
of a field may depend on the intensity of the field. The experimental
attempts to find the physical consequencies of the changed Fermi or 
Bose statistics of different types have been done in~[18--20].

As we have shown, the dynamical equation for quantum observable does
not fix commutation relations and Hamiltonian in complete analogy with
classical dynamical equations which may be described using different
pairs\,:\, Hamiltonian--Poisson brackets. It means that the same quantum 
evolution equation for a physical observable may be compatible with
different pairs\,:\, quantum Hamiltonian--commutation relations for the 
observable. Physical consequence of the
found result for quantum observable evolution is the following. If one 
wants to find out commutation relation of the physical observables
using only known quantum dynamics for these observables, the quantum 
dynamics contains incomplete information to solve the problem and it
yields only some constraints for the commutation relations.

The complete information to fix the commutation relations is contained
in additional physical conditions of measurement procedure for 
measuring the observable. Thus, measuring the quadrature components
gives experimentally established uncertainty relation. The result
found in the work shows that the same quantum dynamics may exist both
for observables which additionally satisfy the uncertainty relation 
for the oscillator quadratures and for the observables which do not
satisfy the uncertainty relation since the commutation relations for
these observables are different from standard position and momentum
commutation relations. 

Thus, with the quantum mechanics formalism are compatible different kinds 
of the same vibrational motion. For standard one, the quadratures
satisfy in addition the uncertainty inequality. For others, they may
not satisfy this inequality. Only physical nature of the vibrating 
quadratures which is determined by measurements distinguishes the two 
equally evolving, in the form of vibrations, quantum observables.
This observation answers the question posed by Wigner: up to what 
extent the quantum equations of motion determine the quantum mechanics
commutation relations? We have shown that there is an essential 
ambiguity in chosing the commutation relations which may be reduced
only taking into account the measurement procedure for the observables.

\section*{Acknowledgements}

\noindent

V. I. M. thanks Osservatorio Astronomico di Capodimonte for hospitality.

\end{document}